\documentstyle[prb,aps,12pt]{revtex}
\begin{document}
\title{Electron Double Refraction in Hybrid Systems with Rashba Spin-Orbit 
Coupling}
\author{V.Marigliano Ramaglia, V.Cataudella, G.De Filippis, C.A.Perroni, F.Ventriglia}
\address{{\sl INFM and Dipartimento di Scienze Fisiche,}\\
{\sl Universit\`{a} degli Studi di Napoli ``Federico II''}\\
{\sl Complesso Universitario Monte Sant'Angelo,}\\
{\sl Via Cintia, 80126 Napoli, Italy}}
\date{\today}
\maketitle

\begin{abstract}
We study the scattering of an electron in a definite state of spin at an
interface of an hybrid system with a Rashba spin-orbit coupling on one side. 
Out of the normal incidence the double refraction
phenomenon appears, with one or two limit angles for the total reflection.
We show that this double refraction gives rise to a spin-dependent conductance 
of a Quantum Point Contact separating a ferromagnet and a two dimensional 
electron gas. The birefringence allows the spin filtering with a single 
interface.
\end{abstract}

\newpage

\section{Introduction}

The most popular spin-filter device,\cite{Ziese,DasSarma}
based on Rashba Effect, \cite{Rashba} has been proposed by Datta and Das \cite
{Datta} as the electronic analog of an electro-optic modulator. The idea is
to design a structure in which the spin behaves as the polarization of the
light. When the light traverses an electro-optic material, two perpendicular
polarizations accumulate different phases shifts and, when the beam emerges
into the analyzer, the two components interfere with each other. Polarizing 
the light
at 45${{}^{\circ }}$ in the plane $\left( y,z\right) ,$ orthogonal to the
direction propagation $x$, the output power collected by the analyzer, that
is oriented in the same way of the input filter, is given by 
\begin{equation}
P_0=\frac 14\left| \,\,\,\,\underline{\overline{%
\begin{array}{ll}
1 & 1
\end{array}
}}\,\,\,\left| 
\begin{array}{l}
e^{ik_1L} \\ 
e^{ik_2L}
\end{array}
\right| \right| ^2=\cos ^2\frac{\left( k_1-k_2\right) L}2.  \label{ras1}
\end{equation}
A gate voltage controls the differential phase shift $\left( k_1-k_2\right)
L $ and allows for a modulation of the output.

In the spintronic analog the role of the electro-optic material is played by
a strip of 2DEG (two dimensional electron gas) in which magnetized contacts
inject and collect electrons with a specific spin orientation. The
feasibility of spin injection at a ferromagnet--semiconductor interface has
been experimentally demonstrated.\cite{Hammar}

An electrical field ${\cal E}$ (in $y$ direction) acts on the strip and the
electrons move along $x$ direction. We suppose that the motion happens in a
nanostructure at low temperature so that the electronic phase coherence is
maintained. The velocities of the charge carriers are of the order $10^8$%
m/sec or larger and a magnetic field (directed in $-z$ direction) appears in
the rest reference frame of the charges. This kind of spin-orbit effect is 
known as Rashba effect. The Hamiltonian spin-orbit term has the form 
\begin{equation}
H_{SO}=\frac{g\left| e\right| }{m^2c^2}\left( \vec{p}\times {\cal \vec{E}}%
\right) \cdot \vec{s}=\frac{g\left| e\right| }{m^2c^2}\left( \vec{s}\times 
\vec{p}\right) \cdot {\cal \vec{E}},  \label{ras2}
\end{equation}
where $g$ is the giromagnetic ratio of the spin $\vec{s}$ and $m$ stands for
the effective mass of the electron charge $-\left| e\right|$. Introducing
the Pauli matrices $\sigma _x,\sigma _y, \sigma _z$, we get 
\[
H_{so}=\frac \eta \hbar \left( \sigma _zp_x-\sigma _xp_z\right)
\]
where $\eta =\left(g\left| e\right| \hbar ^2/2m^2c^2\right) {\cal E}$. 

The motion in $x$ direction is considered taking momentum eigenvalues $p_z=0$
and $p_x=\hbar k^{\prime }$. In this subspace the spin component in $z$
direction is a motion constant and the energy eigenvalues of 
spin up (+) and spin down (--) states are respectively: 
\begin{equation}
E_{\pm }=\frac{\hbar ^2}{2m}k^{\prime \,\,2}\pm \eta k^{\prime }.
\label{ras3}
\end{equation}
For the InGaAs/InAlAs heterostructure the spin-orbit parameter $\eta $ was
estimated to be $\sim 3.9\times 10^{-12}$ eV m.\cite{Datta} For positive
values of the energy we introduce the vector $k$ such that $E_{\pm }=\hbar
^2k^2/2m$. The two spin orientations have two different values of the
momentum 
\begin{equation}
k^{\prime }=k_{\pm }=\sqrt{k^2+\frac{m^2\eta ^2}{\hbar ^4}}\mp \frac{m\eta }{%
\hbar ^2}.  \label{ras4}
\end{equation}

At high values of the energy $(E\gg m\eta ^2/\hbar ^2)$ the two values of $%
k^{\prime }$ have a difference of $2m\eta /\hbar ^2$. Preparing the electron
in the state $\frac 1{\sqrt{2}}\left| 
\begin{array}{l}
1 \\ 
1
\end{array}
\right| $, when it traverses a distance $L$, 
the probability that it can be found
again in this state is given by: 
\begin{equation}
P=\frac 14\left| \,\,\,\underline{\overline{%
\begin{array}{ll}
1 & 1
\end{array}
}}\,\,\left| 
\begin{array}{l}
e^{ik_{+}L} \\ 
e^{ik_{-}L}
\end{array}
\right| \right| ^2=\cos ^2\frac{2m\eta L}{\hbar ^2}.  \label{ras5}
\end{equation}

The Datta and Das ideas have inspired some investigation on spintronic
devices that exhibit spin-valves effects.\cite{Gardelis,Koga}
Recently it has been shown that an oscillatory spin-filtering due to a
spin-dependent conductance can not be obtained by a single interface with 
the Rashba coupling on one side,\cite{Zulicke,Molenkamp} unlikely what 
previously stated.\cite{Grundler} This issue stems from the
boundary conditions that
guarantee the continuity of the probability current density perpendicular to
the interface. These conditions imply the same transmission amplitudes 
for spin up and 
spin down\cite{Zulicke,Molenkamp} and so the effect is absent.
The aim of this paper is to show that the double refraction arising when the 
electron incidence is out of normal mixes the in and out spin states allowing 
for an oscillatory behavior of the conductance with respect to ingoing 
spin polarization. 

\section{Spin scattering against interface}

In this section we will study an hybrid system with an $H_{so}$ coupling 
described by the Hamiltonian:   
\begin{equation}
H=\vec{p}\frac 1{2m\left( x\right) }\vec{p}+\frac{\eta \left( x\right) }\hbar
\left( \sigma _zp_x-\sigma _xp_z\right) -i\sigma _z\frac 12\frac{\partial
\eta \left( x\right) }{\partial x}+U\delta \left( x\right).  \label{ras6}
\end{equation} 
The spatial variation of the
spin-orbit coupling $\eta $ and of the effective mass $m$ on passing from one
side of the interfaces to the other are taken into account in such a way to 
ensure Hamiltonian hermiticity. The kinetic energy and the spin-orbit 
$H_{so}$ contain the momentum operator and have been symmetrized in 
Eq.(\ref{ras6}).\cite{Zulicke} 

We assume that the parameters are piecewise constant 
\begin{eqnarray}
m\left( x\right) &=&m_F\vartheta \left( -x\right) +m_S\vartheta \left(
x\right)  \label{ras7} \\
\eta \left( x\right) &=&\eta \vartheta \left( x\right),  \nonumber
\end{eqnarray}
where $\vartheta \left( x\right) $ is the step function. We have added a term 
$U\delta \left( x\right) $ to control the transparency of the interface. 
The spinor eigenstate of $H,\psi $ is continuous while its derivative has a
discontinuity fixed by the strength of the Dirac delta in $x=0$:  
\begin{eqnarray}
\psi \left( 0+\right) &=&\psi \left( 0-\right)  \label{ras8} \\
\frac{\partial \psi \left( 0+\right) }{\partial x}-\mu \frac{\partial \psi
\left( 0-\right) }{\partial x} &=&\left( u-ik_0\sigma _z\right) \psi \left(
0\right)  \nonumber
\end{eqnarray}
where $\mu =m_S/m_F$, $u=2mU/\hbar ^2$ and $k_0=m\eta /\hbar ^2.$

The free 2DEG with Rashba term occupying the whole $x-z$ plane has the
spinors:  
\begin{eqnarray}
\psi _{+} &=&\exp i\left( k_1x+k_2z\right) \left( 
\begin{array}{l}
\sin \theta \\ 
-\cos \theta
\end{array}
\right)  \label{ras9} \\
\psi _{-} &=&\exp i\left( k_1x+k_2z\right) \left( 
\begin{array}{l}
\cos \theta \\ 
\sin \theta
\end{array}
\right)  \nonumber
\end{eqnarray}
as energy eigenstates of the eigenvalues:  
\begin{equation}
E_{\pm }=\frac{\hbar ^2}{2m_S}\left( k_1^2+k_2^2\right) \pm \eta \sqrt{%
k_1^2+k_2^2},  \label{ras10}
\end{equation}
where $\hbar k_1$ and $\hbar k_2$ are, respectively, the eigenvalues of the
momentum components $p_x$ and $p_z$, and 
\begin{equation}
\theta =\arctan \left[ \frac{k_1}{k_2}+\sqrt{\frac{k_1^2}{k_2^2}+1}\right] .
\label{ras11}
\end{equation}
If $k^{\prime }=\sqrt{k_1^2+k_2^2}$ is the modulus of the momentum and $
\phi $ gives the direction of the in-plane motion ( $k_1/k_2=\cot \phi $ ), 
then 
\begin{equation}
\theta =\frac \pi 2-\frac \phi 2  \label{ras12}
\end{equation}
and 
\begin{equation}
E_{\pm }=\frac{\hbar ^2}{2m_S}\left( k^{\prime \,\,2}\pm 2k_0k^{\prime
}\right) .  \label{ras13}
\end{equation}

We note that the two spinors $\psi _{+}$ and $\psi _{-}$ are one the time
reversed of the other.\cite{Mirales} The time reversal operator $\widehat{T}$%
\[
\widehat{T}\left( 
\begin{array}{l}
\psi _1 \\ 
\psi _2
\end{array}
\right) =\left( 
\begin{array}{l}
\psi _2^{*} \\ 
-\psi _1^{*}
\end{array}
\right) 
\]
commutes with $H_{SO}$: $\left[ \widehat{T},H_{SO}\right] =0$. The degeneracy
with respect to $k^{\prime }$ is lifted but the Rashba term is not able
to produce a spontaneous spin polarization of the electron states: any given
energy value fixes two different values of the modulus $k^{\prime }$, leaving 
undetermined the spin
polarization. However we emphasize that the {\em direction} of the wave
vector $\vec{k}^{\prime }$ fixes the spin polarization as the equation (\ref
{ras12}) shows. If we choose the in-plane motion direction then we put the
electron in a definite spin polarization state. If $\vec{k}^{\prime }$ is
along $x$ direction then $\phi =0$ and $\psi _{+},\psi _{-}$ are the up and
down spins in $z$ direction. Including the spin-orbit interaction in the
Hamiltonian the double group is the new space group \cite{Madelung} and a
space rotation of $4\pi $ is needed to achieve the same spinor.

Let be 
\[
E=\frac{\hbar ^2}{2m_F}k^2 
\]
the electron energy in the ferromagnet. When the electron goes into the
2DEG region its wave vector $k^{\prime }$ becomes 
\begin{equation}
k^{\prime }=\sqrt{\mu k^2+k_0^2}\mp k_0=k_{\pm },  \label{ras14}
\end{equation}
where the index $\pm $ refers to the two branches with the same energy $E$%
\[
E_{\pm }=\frac{\hbar ^2}{2m_S}\left( k_{\pm }^2\pm 2k_0k_{\pm }\right) . 
\]

The hybrid system is invariant for translation along $z$ direction and the
component $p_z$ of the momentum is conserved. If $\alpha $ is the angle of $%
\vec{k}_{+}$ with the $x$ axis, $\beta $ and $\gamma $ the corresponding
angles of $\vec{k}_{-}$ and $\vec{k}$, respectively, the momentum
conservation implies that 
\begin{equation}
k_{+}\sin \alpha =k_{-}\sin \beta =k\sin \gamma.  \label{ras15}
\end{equation}
The figure 1 shows the output angles $\alpha $ and $\beta$. Only when the
incidence is normal, i.e. $\gamma =0$, the electron moves in the 2DEG in the
same direction with $\alpha =\beta =0$. In all other cases, i.e. $\gamma >0$, the
outgoing states $+$ and $-$ go along two different directions. 
This phenomenon is analogous to the double refraction that appears in biaxial
crystals.\cite{Stone} Again the spin of the electrons behaves as the
polarization of the light. We remember that the 
birefringence arises when the characteristics
of electromagnetic propagation depend on the directions of propagation and
polarization of the wave.

The mode $+$ has the limit angle 
\begin{equation}
\gamma _0=\arcsin \frac{k_{+}}k,   \label{ras16}
\end{equation}
so that for $\gamma >\gamma _0$ this mode is totally reflected: it vanishes
exponentially for $x>0$. Here and in the following we take $0<\mu <1$
because the effective mass in the 2DEG is less than in the ferromagnet. When 
$k/k_0<2/\left( 1-\mu \right) $ the mode $-$ is always transmitted up to
grazing incidence at $\gamma =\pi /2$. Increasing the kinetic energy with
respect to spin-orbit coupling, when $k/k_0>2/\left( 1-\mu \right)$, a second
limit angle appears: 
\begin{equation}
\gamma _1=\arcsin \frac{k_{-}}{k} > \gamma _0  \label{ras17}
\end{equation}
and, for $\gamma >\gamma _1$, we have the total reflection (both the modes
vanish for $x>0$). When the strength of spin-orbit coupling goes to zero, $
\gamma _0$ and $\gamma _1$ tend to the common limit $\arcsin \sqrt{\mu }$. 

Lighter the effective mass within the 2DEG is, nearer to normal
incidence the propagation directions $\alpha $ and $\beta $ into Rashba
region $x>0$ are. 
The figure 2 shows the limit angles as a function of $k/k_0$.

The incoming spinor 
\begin{equation}
\psi _i=\exp \left( ik\left( x\cos \gamma +z\sin \gamma \right) \right)
\left( 
\begin{array}{l}
\cos \delta \\ 
\sin \delta
\end{array}
\right)  \label{ras18}
\end{equation}
is reflected at the interface $x=0$ as 
\begin{equation}
\psi _r=\exp \left( ik\left( -x\cos \gamma +z\sin \gamma \right) \right)
\left( 
\begin{array}{l}
r_{+} \\ 
r_{-}
\end{array}
\right)  \label{ras19}
\end{equation}
and transmitted at $x>0$ in both the modes $+$ and $-$ as 
\begin{eqnarray}
\psi _t &=&t_{+}\exp \left( ik_{+}\left( x\cos \alpha +z\sin \alpha \right)
\right) \left( 
\begin{array}{c}
\cos \alpha /2 \\ 
\sin \alpha /2
\end{array}
\right) +  \label{ras20} \\
&&\dot{t}_{-}\exp \left( ik_{-}\left( x\cos \beta +z\sin \beta \right)
\right) \left( 
\begin{array}{c}
-\sin \beta /2 \\ 
\cos \beta /2
\end{array}
\right),  \nonumber
\end{eqnarray}
where $\delta $ fixes the spin polarization of electron within the
ferromagnet. When $\gamma =0$ a spin up (along $z$ direction) goes in the
mode $+,$ while the spin down propagates in the mode $-$ at $x>0.$ In this
case Z\"{u}like {\it et al.}\cite{Zulicke} and Molenkamp {\it et al.} \cite{Molenkamp}
have shown that $t_{+}=t_{-}$ and the interface is not able to filter the
spin. We note that out of the normal incidence with $\gamma >0$ the
scattering changes the spin polarization. The transmitted amplitudes $%
t_{+},t_{-}$ and the reflected ones $r_{+},r_{-}$ are determined by the
boundary conditions (\ref{ras8}) as functions of $k,k_0,u,\delta $ and $%
\gamma $. Solving the system 
\[
t_{+}\cos \frac \alpha 2-t_{-}\sin \frac \beta 2=\cos \delta +r_{+} 
\]
\[
t_{+}\sin \frac \alpha 2+t_{-}\cos \frac \beta 2=\sin \delta +r_{-} 
\]
\[
k_{+}t_{+}\cos \alpha \cos \frac \alpha 2-k_{-}t_{-}\cos \beta \sin \frac %
\beta 2-\mu k\cos \gamma \left( \cos \delta -r_{+}\right) = 
\]
\begin{equation}
-\left( k_0+iu\right) \left( \cos \delta +r_{+}\right)  \label{ras21}
\end{equation}
\[
k_{+}t_{+}\cos \alpha \sin \frac \alpha 2+k_{-}t_{-}\cos \beta \cos \frac %
\beta 2-\mu k\cos \gamma \left( \sin \delta -r_{-}\right) = 
\]
\[
\left( k_0-iu\right) \left( \sin \delta +r_{-}\right) . 
\]
We get 
\[
r_{+}=(C_{+}A_{--}\cos \alpha /2-C_{-}A_{+-}\sin \beta /2)/D 
\]
\[
r_{-}=(C_{-}A_{++}\cos \beta /2-C_{+}A_{-+}\sin \alpha /2)/D 
\]
\begin{equation}
t_{+}=\left[ \left( \cos \delta +r_{+}\right) \cos \beta /2+\left( \sin
\delta +r_{-}\right) \sin \beta /2\right] /\cos \frac{\alpha -\beta }2
\label{ras22}
\end{equation}
\[
t_{-}=\left[ \left( \sin \delta +r_{-}\right) \cos \alpha /2-\left( \cos
\delta +r_{+}\right) \sin \alpha /2\right] /\cos \frac{\alpha -\beta }2 
\]
with 
\begin{eqnarray*}
A_{++} &=&k_{+}\cos \alpha +\mu k\cos \gamma +k_0+iu \\
A_{+-} &=&k_{+}\cos \alpha +\mu k\cos \gamma -k_0+iu \\
A_{-+} &=&-k_{-}\cos \beta -\mu k\cos \gamma -k_0-iu \\
A_{--} &=&k_{-}\cos \beta +\mu k\cos \gamma -k_0+iu
\end{eqnarray*}
\begin{eqnarray*}
C_{+} &=&\left( -k_{+}\cos \alpha +\mu k\cos \gamma -k_0-iu\right) \cos
\delta \cos \frac \beta 2+ \\
&&\left( -k_{+}\cos \alpha +\mu k\cos \gamma +k_0-iu\right) \sin \delta \sin 
\frac \beta 2
\end{eqnarray*}
\begin{eqnarray*}
C_{-} &=&\left( k_{-}\cos \beta -\mu k\cos \gamma +k_0+iu\right) \cos \delta
\sin \frac \alpha 2+ \\
&&\left( -k_{-}\cos \beta +\mu k\cos \gamma +k_0-iu\right) \sin \delta \cos 
\frac \alpha 2
\end{eqnarray*}
\[
D=A_{++}A_{--}\cos \frac \beta 2\cos \frac \alpha 2-A_{+-}A_{-+}\sin \frac %
\beta 2\sin \frac \alpha 2. 
\]
We note that when $\gamma >\gamma _0$ then 
\[
\sin \alpha =\frac k{k_{+}}\sin \gamma >1,  
\]
whose solution in $\alpha $ is 
\[
\alpha =\frac \pi 2+i\alpha ^{\prime }\,\,\,;\,\,\,\sin \alpha =\cosh \alpha
^{\prime }\,;\,\,\,\cos \alpha =-i\sinh \alpha ^{\prime }. 
\]
The mode $+$ becomes a vanishing wave along $x$ axis:  
\[
t_{+}\exp \left( -k_{+}x\sinh \alpha ^{\prime }\right) \exp \left(
ik_{+}z\cosh \alpha ^{\prime }\right) \left( 
\begin{array}{l}
\cos \left( \pi /4+i\alpha ^{\prime }/2\right) \\ 
\sin \left( \pi /4+i\alpha ^{\prime }/2\right)
\end{array}
\right) . 
\]
When $\gamma >\gamma _1,$ $\beta =\pi /2+i\beta ^{\prime }$ and both the
modes are damped within the 2DEG: the incident wave is totally reflected.

At normal incidence 
\[
\gamma =\alpha =\beta =0 
\]
and 
\begin{eqnarray*}
t_{+} &=&\frac{2\mu k\cos \delta }{k_{+}+k_0+iu+\mu k} \\
t_{-} &=&\frac{2\mu k\sin \delta }{k_{.-}-k_0+iu+\mu k}. 
\end{eqnarray*}
Since
\begin{equation}
k_{+}+k_0=k_{.-}-k_0=\sqrt{\mu k^2+k_0^2}  \label{ras22a}
\end{equation}
the transmitted spinor is: 

\begin{eqnarray*}
\psi _t &=&\frac{2\mu k}{\sqrt{\mu k^2+k_0^2}iu+\mu k}\left(
e^{ik_{+}x}\left( 
\begin{array}{l}
\cos \delta \\ 
0
\end{array}
\right) \right. \\
&&\left. +e^{ik_{-}x}\left( 
\begin{array}{l}
0 \\ 
\sin \delta
\end{array}
\right) \right)
\end{eqnarray*}

and the interference between the modes $+$ and $-$ at the interface in $x=0$
is lost. If $\delta =\pi /4$, projecting $\psi _t$ on the input spinor, we get 
\[
\left| \psi _t^{\dagger }\times \left( 
\begin{array}{l}
1/\sqrt{2} \\ 
1/\sqrt{2}
\end{array}
\right) \right| ^2 \propto \cos ^2\frac{k_{+}-k_{-}}2x 
\]
that is the Datta and Das \cite{Datta} modulation factor. However an
analyzer, i.e. a second interface, is needed to have a spin dependent
transmission.

The square moduli of the transmitted amplitudes $\left| t_{\pm }\left(
\delta \right) \right| ^2$ are shown in fig.3 when $\gamma $ is between $0$
and $\pi /2$. We see how $\left| t_{\pm }\left( 0\right) \right| ^2$ and $%
\left| t_{\pm }\left( \pi /2\right) \right| ^2$, and $\left| t_{\pm }\left(
\pi /4\right) \right| ^2$ and $\left| t_{\pm }\left( 3\pi /4\right) \right|
^2$ too, start from the same value for $\gamma =0$ but are different when
the incidence angle increases towards $\pi /2$. The derivatives of $\left|
t_{\pm }\left( \delta \right) \right| ^2$ jumps at $\gamma _0$ and
then at $\gamma _1$, when the character of the mode propagation changes. The
traversing of the interface changes the spin polarization when $\gamma >0$.

We get the transmission coefficient $T$ from the probability current density 
\begin{equation}
\begin{array}{ll}
\vec{j}={\Re}\left\{ \psi ^{\dag }\vec{p}\psi \right\} & ;\,\,x<0 \\ 
\vec{j}={\Re}\left\{ \psi ^{\dag }\left( \vec{p}+\hbar k_0\cdot \widehat{%
y}\times \vec{\sigma}\right) \psi \right\} & ;\,\,x>0
\end{array}
\label{ras23}
\end{equation}
whose $x-$component is 
\begin{eqnarray}
j_{xr} &=&\hbar k\cos \gamma \left( 1-\left| r_{+}\right| ^2-\left|
r_{-}\right| ^2\right) /m_F\,\,  \label{ras24} \\
\text{for}\,\,x &<&0\,\,\, \text{and}\  \nonumber \\
j_{xl} &=&\left[ \hbar \left( k_{+}+k_0\right) \cos \alpha \cdot \left|
t_{+}\right| ^2+\right.  \nonumber \\
&&\left. \left( k_{-}-k_0\right) \cos \beta \cdot \left| t_{-}\right|
^2\right] /m_S  \nonumber \\
\text{for }\,\,x &>&0.  \nonumber
\end{eqnarray}
The boundary conditions (\ref{ras8}) assure the continuity of $j_x$ as can be
verified by a straightforward calculation of eqs.(\ref{ras24}). When $\gamma
<\gamma _0$ both the modes propagate in $x>0,$ the only $-$ mode remains
when $\gamma _0<\gamma <\gamma _1$. 

The transmission coefficient is the ratio of 
$j_{xr}$ with the incident flux $j_i=\hbar k\cos \gamma /m_F$, $T=j_{xr}/
j_i$, while the reflection coefficient is 
$R=\left( j_{i} - j_{xr} \right) /j_{i} $: 
\begin{eqnarray}
T\left( \delta ,\gamma \right) &=&\left( \left( k_{+}+k_0\right) \cos \alpha
\cdot \left| t_{+}\right| ^2\vartheta \left( \gamma _0-\gamma \right)
+\right.  \label{ras25} \\
&&\left. \left( k_{-}-k_0\right) \cos \beta \cdot \left| t_{-}\right|
^2\vartheta \left( \gamma _1-\gamma \right) \right) /\mu k\cos \gamma 
\nonumber \\
R\left( \delta ,\gamma \right) &=&\left| r_{+}\right| ^2+\left| r_{-}\right|
^2  \nonumber
\end{eqnarray}
and when $\gamma $ overcomes $\gamma _1$ , $T\left( \delta ,\gamma \right)
=0 $ and $R\left( \delta ,\gamma \right) =1$. The flux is conserved because
in all the cases 
\[
T\left( \delta ,\gamma \right) +R\left( \delta ,\gamma \right) =1. 
\]

The transmission coefficient as a function of $\gamma $ has a first higher
step up to $\gamma _0$ followed by a lower step that ends in $\gamma _1$. 
The fig.4 shows how the shapes and the heights of the two steps vary with the
spin polarization angle $\delta$. At low values of $\mu $, that is the
electrons in 2DEG are light, the propagation in the $x>0$ region happens at
angles nearer to the normal incidence. At equal masses $(\mu =1)$ the
passage is allowed up to grazing incidence and the steps appear more
squared. We note that the second step tends to disappear around $\delta =\pi
/4$ and has the maximum height around $\delta =3\pi /4.$ The fig.5 refers to
the case of an higher Fermi wave vector $k.$ Obviously when $%
k/k_0\rightarrow \infty $, $T=1$ for $\gamma $ from $0$ to $\pi /2$ but the
second step is again visible for $k$ greater then $k_0$ of two magnitude
orders.

\section{Quantum Point Contact Conductance}

The previously described double refraction can affect the conductance of a 
ballistic quantum point contact. 

Let a constriction of width $W$ separate the ferromagnet and the 2DEG 
that behave as two perfect reservoirs at the Fermi energy:  
\[
E_F=\frac{\hbar ^2k^2}{m_F}=E_{\pm }=\frac{\hbar ^2}{2m_S}\left( k^{\prime
\,\,2}\pm 2k_0k^{\prime }\right). 
\]
The electron motion within the hybrid system is assumed to be ballistic;
that is the electronic mean free path is much longer than the size $W$ of 
the point contact. The Landauer-B\"{u}ttiker formalism applies \cite{Buttiker,Beenakker}.The conductance $G$ at zero temperature is given by 
\begin{equation}
G=\frac{e^2}h\sum_iT_i,  \label{ras26}
\end{equation}
where $T_i$ are the transmission coefficients for all the open channels $i\,$%
between the two reservoirs at the energy $E_F$ .In our case the index $i$
represents the incidence angle $\gamma .$

A sketch of the point contact can be found in fig.6a). The 2D Fermi circle
in k-space appears in fig.6b) and only the states on its edge can carry
current at zero temperature. As we have shown before, the current is
transported through the point contact by the states belonging to the arch
from $-\gamma _1\,$ to $\gamma _1$ on the Fermi circle. 

Quantum mechanically,
the current through the point contact is equipartitioned among the $1D$
sub-bands, or transverse modes, in the constriction. The gap along $k_z$ axis
between two consecutive sub-bands can be estimated of the order of $\pi /W$
(this is exactly the result for a square well lateral confining potential of
width $W$). The number of states contained in the element of arch $d\gamma $
is then $kd\gamma /\left( \pi /W\right) .$ The equation (\ref{ras26})
implies that hybrid system conductance $G$ is 
\begin{equation}
G=\frac{e^2}h\int_{-\gamma _1}^{\gamma _1}T\left( \delta ,\gamma \right) 
\frac{kWd\gamma }\pi =\frac{e^2kW}h{\cal G}\left( \delta \right)
\label{ras27}
\end{equation}
with 
\begin{equation}
{\cal G}\left( \delta \right) =\frac 1\pi \int_{-\gamma _1}^{\gamma
_1}T\left( \delta ,\gamma \right) d\gamma .  \label{ras28}
\end{equation}

An exhaustive discussion about this approach can be found in references 15 and 16. We note that the restriction to the normal
incidence $\gamma =0$ gives 
\[
{\cal G}\left( \delta \right) =\frac{T\left( \delta ,0\right) }\pi 
\]
that is the Sharvin resistance formula \cite{Sharvin} used by Grundler 
\cite{Grundler}
but that is independent on the spin polarization angle $\delta$. 

The fig.7 shows ${\cal G}\left( \delta \right) $ for $\delta $ between 0 and 
$\pi$. The oscillatory behavior of the conductance allows the spin
filtering with a single interface. This effect is a direct consequence of
the double refraction at the interface that changes the spin state when the
electron enters the region where the spin-orbit Rashba coupling
works. At normal incidence the electron pass into 2DEG conserving the spin
state. When a lateral confining potential is imposed to the electron gas the
Q1DEG has sub-bands for which the free electron property (\ref{ras22a}) is no
more valid, although the time reversal symmetry leaves the degeneracy of
states with opposite value of $k_1$. The case of a parabolic confining
potential has been studied by Governale {\it et al.} \cite{Governale} that
estimate the deformation of sub-bands and the lateral spin density. The
ballistic spin-transport properties of a quasi-one-dimensional wire with a
spin-orbit Rashba interaction in a finite piece of it have been studied with
a numerical tight binding model by Mirales {\it et al.}\cite{Mirales} They find
a spin-conductance modulation.

An estimation of the strength of Rashba interaction on the conductance 
is given by the ratio: 
\[\Delta {\cal G}/{\cal G}=
\frac{{\cal G}\left( 3\pi /4\right) -{\cal G}\left( \pi /4\right) }{{\cal G}%
\left( 0\right) } 
\]
reported in Fig.8.  
That is roughly the maximum relative variation of the conductance against $%
k/k_0$. We note that for $k/k_0=100$, $\Delta {\cal G}/{\cal G}$ is of the
order of ten per cent. We think that such a value could be detected 
experimentally in a quantum point contact.   

\section{Conclusions}
In this paper transmission across a ferromagnet/2DEG has been studied. An
electron in a definite state of spin undergoes a double refraction
traversing the interface analogously to what happens to the polarized light
impinging the surface of a biaxial crystal. We have shown that the correct
boundary conditions give rise to a spin-dependent transmission coefficient
and that the normal incidence is a special case for which the dependence on 
spin is lost. The spin filtering occurs when the electron hits the interface
in a direction out of the normal. The conductance of a point contact at the
interface in ballistic transport regime within the Landauer-B\"{u}ttiker
formalism has been calculated. We have shown 
that the conductance has an oscillatory
behavior with the polarization angle of the spin.

We gratefully acknowledge M.Governale for helpful suggestions about the
correct way to impose the boundary conditions at the interface.

\begin{center}
{\bf Figure Captions}
\end{center}

\begin{enumerate}
\item[Fig:1]  The vectors $\vec{k}_{+},$ $\vec{k}_{-},$ $\vec{k}$ in $k-$%
space and the angles $\alpha ,\beta $ and $\gamma $ that they form with $x$
direction normal to the interface. The two circles are the lines at the
constant energy $\hbar ^2k^2/2m_F.$

\item[Fig.2]  The limit angles $\gamma _0$ of $+$ mode (dashed line) and $%
\gamma _1$ of $-$ mode (full line) for the indicated values of mass ratio $%
\mu $ as functions of $k/k_0$. For $\gamma $ above $\gamma _1$ the total
reflection occurs.

\item[Fig.3]  The squared moduli of the transmitted amplitudes for two
couples of orthogonal spin polarizations. The cusps sign the passage through
the limit angles.

\item[Fig.4]  The two steps of the transmission coefficient $T$. The second
step tends to disappear for $\delta =\pi /4$ and to have the same height of
the first when $\delta =3\pi /4$.

\item[Fig.5]  The steps of $T$ at the higher value of $k/k_0=100.$ In the
limit $k/k_0\rightarrow \infty ,$ $T=1$ for $0<\gamma <\pi /2.$

\item[Fig6.]  a)The sketch of the point contact\newline
b) The Fermi circle in $k-$space. The thick arch indicates the states that
carry current into the point contact.

\item[Fig.7]  The conductance ${\cal G}$ as a function of the polarization
spin angle $\delta$.

\item[Fig:8]  The relative variation of the conductance $\Delta {\cal G}/
{\cal G}$ against $k/k_0$.
\end{enumerate}

\end{document}